\documentclass[letterpaper, 10 pt, conference]{ieeeconf} 
\IEEEoverridecommandlockouts 
\usepackage{graphicx} % Required for inserting images
\usepackage{booktabs} % For professional quality lines
\usepackage{pifont}   % For tick and cross symbols
\usepackage{amssymb}  % For the question mark symbol
\usepackage{listings}
\usepackage{xcolor}
\usepackage{float}
\usepackage{geometry}
\usepackage{glossaries}

 \usepackage{xurl}   % TO DO: Aktuell brechen die Links die Linewidth. Mit diesem Package sind sie aber nicht mehr klickbar (werden abgehakt).

\newgeometry{top=1in, bottom=0.75in, left=0.75in, right=0.75in}

\title{\LARGE \bf
Sense4HRI: A ROS 2 HRI Framework for Physiological Sensor Integration and Synchronized Logging
}

\author{Manuel Scheibl$^{1}$, Julian Leichert$^{1,2}$, Sinem Görmez$^{1}$, Britta Wrede$^{1}$
\thanks{$^{1}$Medical Assistance Systems, Medical School OWL, Bielefeld University, 33501 Bielefeld, Germany}
\thanks{$^{2}$Center for Cognitive Interaction Technology, CITEC from Bielefeld University, Germany
        {\tt\small manuel.scheib@uni-bielefeld.de, jleichert@uni-bielefeld.de, sinem.goermez@uni-bielefeld.de,
    bwrede3@uni-bielefeld.de}}%
}
\date{February 2026}

\begin{document}

\maketitle
\thispagestyle{empty}
\pagestyle{empty}
\begin{abstract}
Physiological signals are increasingly relevant to estimate the mental states of users in human-robot interaction (HRI), yet ROS 2-based HRI frameworks still lack reusable support to integrate such data streams in a standardized way. Therefore, we propose Sense4HRI, an adapted framework for human–robot interaction in ROS 2 that integrates physiological measurements and derived user-state indicators. The framework is designed to be extensible, allowing the integration of additional physiological sensors, their interpretation, and multimodal fusion to provide a robust assessment of the mental states of users. In addition, it introduces reusable interfaces for timestamped physiological time-series data and supports synchronized logging of physiological signals together with experiment context, enabling interoperable and traceable multimodal analysis within ROS 2-based HRI systems.
\end{abstract}


\section{Introduction}

With robotic systems increasingly moving into everyday and assistive contexts \cite{Martinez_2021}, understanding the user’s current state becomes more important for safe, adaptive, and context-sensitive human–robot interaction (HRI). Physiological signals provide a valuable additional source of information, as they  reveal aspects of the user’s internal state that are not directly observable from behavior alone \cite{Guarnaccia_2026,Savur_2023}. Measures such as heart activity, skin conductance, respiration, and eye-related signals contribute to estimating stress, cognitive load, alertness, and affective arousal, thereby supporting more nuanced interaction and analysis in HRI \cite{Guarnaccia_2026,Savur_2023}.

This potential has become increasingly feasible due to recent advances in wearable sensing. Many physiological sensors have become smaller, less expensive, and easier to integrate into everyday settings \cite{Kaur_2023,Wu_2023}, making continuous and less obtrusive data collection popular. This includes not only common consumer devices such as smartwatches, but also more specialized portable systems, s.a. wireless EEG (electroencephalography) headsets. As a result, multiple physiological modalities can now be acquired in parallel, creating new opportunities for rich physiological Human Digital Twins (HDTs)\cite{lauerschmaltz2024humandigitaltwindefinition} and data-driven interaction analysis.

However, physiological signals are particularly valuable in HRI when they are not considered alone but together with other communicative channels, since human communication is inherently multimodal. Beyond spoken words, meaning is conveyed through vocal cues, nonverbal behavior, and physiological responses \cite{Drijvers_2022}. Accordingly, prior work across disciplines has described communicative meaning as distributed across multiple behavioral and physiological channels that should be considered jointly \cite{Mehu_2015,Udahemuka_2024}. For HRI, incorporating these additional modalities enables more nuanced user-state models \cite{Guarnaccia_2026,Savur_2023}, for example by capturing visual attention, alertness, stress, and task load. This, in turn, supports more context-sensitive robot behavior and richer multimodal analyses of interaction.

To capture and relate these heterogeneous data streams in a unified and reproducible way, an appropriate middleware infrastructure is required. We adopt ROS 2 \cite{ROS2_2026} as the basis for this work because it provides the core mechanisms needed for reproducible multimodal HRI studies. Its standardized publisher–subscriber architecture supports modular integration of distributed components, including robot control, logging, and analysis, while enabling heterogeneous sensors to be incorporated through reusable nodes and packages \cite{Macenski_2022}. In particular, rosbag2 allows multiple time-stamped topic streams to be recorded and replayed within a single artifact, which simplifies debugging and supports reproducible offline analysis \cite{ROS2_2026}. These properties make ROS 2 well suited for physiological sensing in HRI: by publishing physiological measurements as ROS topics, biosignals can be synchronized with interaction context such as robot actions, speech, gaze, and task events, and can be captured together for traceable multimodal analysis and fusion.

However, while ROS 2 provides the technical basis for such integration, currently available examples of physiological sensing in HRI do not yet offer reusable, standardized interfaces (e.g., \cite{jo_toward_2021,quarez_mutual_2025}). Instead, published implementations are often tailored to specific devices or application settings, which limits modularity, extensibility, and reuse across studies. At the same time, ROS4HRI \cite{Youssef_2021} provides an extensible representation framework for humans in complex HRI scenarios, but does not include dedicated support for physiological sensing. In this paper, we address this gap by proposing a ROS 2 framework extension for physiological sensor integration and synchronized logging that is modular, reusable, and interoperable with ROS4HRI-based systems.

\section{Related Work}

\begin{table*}[!t]
    \centering
    \caption{Overview of physiological sensors used in HRI.}
    \label{tab:sensor-evaluation}
    \renewcommand{\arraystretch}{1.15}
    \setlength{\tabcolsep}{6pt}
    \begin{tabular}{@{}l p{5.8cm} p{7.8cm}@{}}
        \toprule
        \textbf{Sensor Type} & \textbf{Measurement} & \textbf{Indicative of}\\ 
        \midrule
        EEG & Multi-Channel Brain Activity \cite{Savur_2023}      & Workload \cite{Savur_2023}; Stress \\
        
        PPG & Variations in blood volume \cite{Savur_2023} &  Anxiety; cognitive effort \cite{Rihet_2024}\\
        
        ECG & Electrical activity of the heart \cite{Savur_2023} &  Mental and Emotional Stress \cite{Shao2021ComparisonHRV, 10731445, Rodriguez_2025}; Physical Effort \cite{Novak_2015}\\
        Skin Conductance  & Electrical conductivity of the skin &  Emotional arousal; mental load \cite{Savur_2023}\\
        
        EMG & Electrical activity generated by skeletal muscles \cite{Savur_2023} & motor control / motor intention \cite{Savur_2023}\\
        
        Eye Tracking & Visual attention and gaze behavior \cite{Savur_2023} & Attention \cite{Savur_2023}; engagement \cite{Kompatsiari_2019}; intention \cite{Zhu_2024}\\
        
        EOG & Electrical potential difference between the front and back of the eye \cite{Abdel_2021} & Eye movement \cite{Abdel_2021}; blink \cite{Ma_2015}; gaze direction \cite{Ma_2015}\\
        
        Pupillometry & Changes in the diameter of the eye's pupil \cite{Hostettler_2023} & Workload \cite{Savur_2023}; emotional arousal \cite{Reuten_2018}; trust \cite{Kret_2015}\\
        
        Respiration & breathing patterns, such as respiratory rate and variability \cite{Novak_2015} & Arousal; workload; effort \cite{Novak_2015}\\
        \bottomrule
    \end{tabular}
\end{table*}

Physiological sensing in HRI involves two complementary perspectives, namely the sensor modalities used to capture physiological processes and the middleware frameworks used to integrate them into robotic systems. Table~\ref{tab:sensor-evaluation} summarizes the physiological sensor modalities commonly used in HRI, their primary measurements, and the higher-level states or processes they have been used to indicate. On the system side, for example, \cite{jo_toward_2021} proposed a multimodal framework to integrate multiple sensors into HRI studies. The authors developed a unified package that encompasses various sensors, with a separate node deployed for each sensor. However, this monolithic approach limits scalability and deployment flexibility, as it requires all sensor dependencies to be resolved before deployment, even when not all sensors are used. In addition, the proposed ROS topic design does not address multi-user HRI scenarios. Furthermore, relying solely on ROS timestamps may introduce synchronization issues when incorporating fusion models across aggregated modalities.

In the medical domain, Quarez et al. proposed the ROS 2 framework MUTUAL to enable agentic systems powered by multimodal data streams in the operating room. It integrates modalities such as RGB-D vision, force sensing, and audio within a cross-platform infrastructure that supports recording and real-time streaming \cite{quarez_mutual_2025}.
However, the contribution focuses primarily on deployment and the application itself, while not addressing topic and message design in detail or providing clear patterns for reusability.

Beyond frameworks directly targeting physiological sensing, ROS 2-based solutions have also been proposed for HRI more broadly. Youssef et al. introduced a standard human representation model for complex HRI scenarios which enables the unique identification of humans along with attributes such as facial expressions, body posture, voice, gaze, engagement, and interaction intent \cite{Youssef_2021}.
However, the framework does not currently provide dedicated interfaces for physiological sensors, even though their use is becoming increasingly common in HRI studies.

While the above work addresses HRI middleware and multimodal interaction more broadly, a smaller line of work has begun to address physiological sensing in robotic systems more directly.

ROS-Neuro \cite{Tonin_2022} provides a modular middleware for neurorobotics that supports acquiring, processing, and decoding signals such as EEG and EMG (electromyography) and connecting them to robotic pipelines. However, its main focus is neurorobotics rather than general HRI. In addition, the version described by Tonin et al. is based on the deprecated ROS 1.

More recently, ROS 2 Healthcare (ROS2HC) \cite{Pena_2024} introduces standardized interfaces for biosignals and derived biometric states, together with sensor drivers, visualization tools, and example pipelines for data analysis and closed-loop control. Among the reviewed ROS 2 approaches, it comes closest in scope to standardized physiological data integration. However, its primary focus lies in healthcare, rehabilitation, and clinical monitoring scenarios. As a result, it does not directly address the need for a lightweight physiological extension tailored to general-purpose HRI experiments.

While several efforts address physiological sensing in robotics and human-centered systems, there is still no broadly adopted ROS 2 solution that combines modularity, reusability, and seamless integration with general-purpose HRI infrastructures. Existing approaches are often tightly coupled, device-specific, or tailored to narrower domains, which limits their extensibility and reuse across applications.

Extensible HRI frameworks for representing the human already exist. However, these frameworks typically do not provide a dedicated and reusable mechanism for integrating physiological sensors in ROS 2-based systems, or they require substantial implementation effort to incorporate such sensing in practice.

This suggests an existing gap between physiological sensing and HRI infrastructure. A lightweight, modular, and reusable ROS 2 framework for physiological sensing that integrates smoothly into general HRI workflows is still missing.

\section{Design}\label{sec:design}

\begin{figure*}
    \centering
    \includegraphics[width=0.95\textwidth]{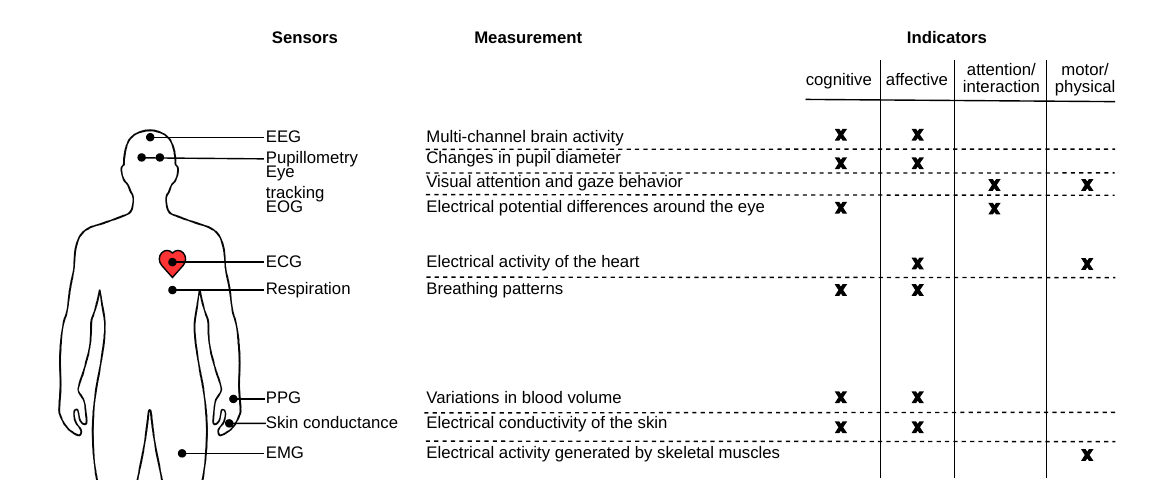}
    \caption{Schematic overview of the physiological sensing pipeline used in this work. Left: exemplary body locations of the considered sensor modalities. Middle: corresponding primary (raw) measurements. Right: high-level indicator categories for which each modality can provide evidence. Crosses indicate a non-exclusive mapping (i.e., a sensor may inform multiple categories and categories may be supported by multiple sensors).}
    \label{fig:physio_sensing}
\end{figure*}

Based on the related work, we developed Sense4HRI\footnote{\url{https://gitlab.ub.uni-bielefeld.de/mas/projects/sense4hri}}, a ROS~2 framework for integrating physiological sensors into existing HRI systems. Its key design goals are modularity, decoupling, and scalable sensor integration, along with clearly defined interfaces that make it easy to add new sensors or develop new applications.  

The framework is also intended to be compatible with existing ROS-based HRI infrastructures, in particular ROS4HRI \cite{Youssef_2021}. ROS4HRI emphasizes clear and reusable interfaces for representing humans and their attributes in complex interaction scenarios, including support for multiple users through unique identifiers. Building on these principles, Sense4HRI organizes physiological sensing as a set of modular ROS~2 packages, with one package per sensor. Each package defines standardized topic and message structures and specifies node responsibilities within the package. This structure allows heterogeneous sensors, including the modalities summarized in Fig.~\ref{fig:physio_sensing}, to be integrated independently, while downstream applications can access both, raw measurements as well as processed signals through a unified interface. The design facilitates the integration of heterogeneous physiological sensors into multimodal HRI experiments while remaining in alignment with the topic and naming conventions set by ROS4HRI.

As illustrated in Fig.~\ref{fig:physio_sensing}, physiological sensing in HRI does not follow a one-to-one mapping between sensing modalities and downstream indicators. The same indicator category may be informed by multiple sensor modalities. To accommodate this non-exclusive relationship, the proposed framework separates generic raw sensor acquisition from sensor-specific interpretation, allowing interpreter logic and processing pipelines to be exchanged independently without modifying the underlying sensor driver. As a result, interpreter logic and processing pipelines can be modified or replaced seperately without any modifications to the underlying sensor driver.

\begin{figure*}
        \centering
    \includegraphics[width=\linewidth]{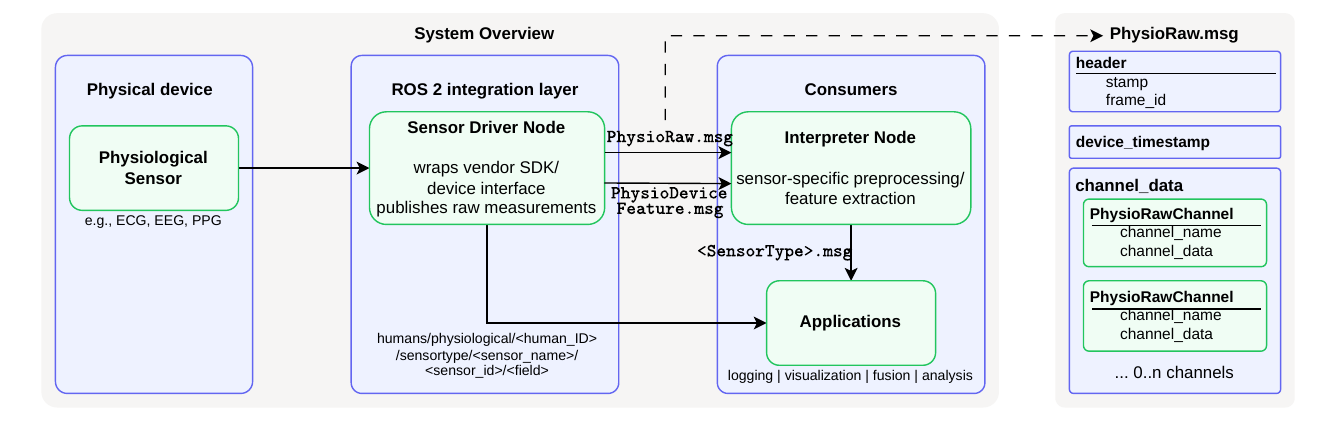}
    \caption{Overview of the proposed ROS 2 architecture for physiological sensing (left) and detailed structure of \texttt{PhysioRaw.msg} (right). A sensor driver node wraps the physiological device and publishes raw measurements as \texttt{PhysioRaw.msg}, which can be consumed directly by applications or processed by an interpreter node for sensor-specific preprocessing and feature extraction. The architecture further supports device-level feature messages (\texttt{PhysioDeviceFeature.msg}) and processed sensor-specific outputs (\texttt{<SensorType>.msg}). The \texttt{PhysioRaw.msg} message comprises ROS metadata (\texttt{header}), a device-side timestamp (\texttt{device\_timestamp}), and an array of \texttt{PhysioRawChannel} elements containing channel names and floating-point sample arrays.}
    \label{fig:Architecture}
\end{figure*}

As shown in Fig.~\ref{fig:Architecture}, each integrated sensor is represented by a dedicated sensor driver node that wraps the corresponding device interface and publishes its raw sensor data on a dedicated ROS topic (e.g., \texttt{<base\_topic>/raw}), which serves as the standard interface for other nodes. These driver nodes handle device-specific connections such as Bluetooth, USB or LabStreamingLayer (LSL) and may also expose additional device-specific features through separate topics (e.g., \texttt{<base\_topic>/device}).

To provide a reusable interface for raw physiological measurements, we define a generic message format for time-series sensor output. As illustrated on the right side of Fig.~\ref{fig:Architecture}, \texttt{PhysioRaw.msg} contains standard ROS metadata (\texttt{header}), a device-side timestamp (\texttt{device\_timestamp}), and an array of named channels. Each channel is represented by a \texttt{PhysioRawChannel} element consisting of a \texttt{channel\_name} and a floating-point sample array. This design allows heterogeneous sensors to expose raw measurements in a unified structure while preserving channel-wise information. 
In addition to this generic raw format, the framework supports a separate sensor message type for features provided by the device.
The topics published by the sensor driver can be consumed directly by applications such as logging, visualization, or multimodal fusion. The may also first be processed by an interpreter node that subscribes to the raw data and publishes preprocessed or derived, sensor-specific features (e.g., \texttt{<base\_topic>/features}). For this purpose we defined multiple sensor specific messages, including common derived features for the sensor type.
For example, an ECG interpreter node would publish an \texttt{ECG.msg} message, which consists of a header, a device timestamp, and multiple derived features such as the RR interval, the number of peaks, Standard Deviation of NN Intervals (SDNN), the heart rate, and other relevant features \cite{Savur_2023}.

This separation of interpretation and device driver keeps the sensor integration modular, which allows application developers to choose between low-level measurements and higher-level representations.

In order to organize the different sensor data streams, we defined a topic structure that extends and aligns with the structure proposed by ROS4HRI.
It is based on the sensor modalities and measurements summarized in Fig.~\ref{fig:physio_sensing} and supports multiple users and sensors.
The topic namespace is designed such that new sensors from different vendors can be integrated without changing the overall structure. Thus, all physiological data is organized in a common topic hierarchy that remains stable across sensor types and vendors:

\begin{center}
  \texttt{/humans/physiological/<human\_id>/\allowbreak<sensor\_type>/<sensor\_id>/<field>}
\end{center}
For example, a topic representing features derived from an EEG headset could be published as:
\begin{center}
    \texttt{/humans/physiological/p1/eeg/\allowbreak headset\_1/features}
\end{center}

This naming scheme makes sensor streams both, distinguishable and extensible, while being applicable in multi-user HRI scenarios.

In addition to the established data streams, each sensor node exposes static hardware metadata via the ROS 2 parameter server at startup. These parameters are loaded from a configuration file passed to the node at launch and can include properties such as sampling frequency, measurement range and unit. Other nodes may query this information through the standard ROS 2 parameter service interface (\texttt{rcl\_interfaces/srv/GetParameters}) in order to adapt processing steps to the characteristics of the underlying hardware at runtime. This mechanism supports reusable downstream components that can reason, not only over the measured values, but also over sensor-specific properties.

Another challenge in integrating multiple physiological sensors is time synchronization across heterogeneous hardware devices. Many sensors provide timestamps based on their own internal clocks. But these do not use a common temporal reference. As a result, absolute timestamps from different devices are generally not directly comparable. To address this issue, the proposed framework includes a device-side timestamp in each incoming sensor message and relates it to ROS time on receipt. This enables the estimation of the offset between device time and ROS time and thereby supports synchronized logging and alignment of multimodal data streams. Such synchronization is particularly important when physiological signals are to be analyzed jointly with other interaction-relevant data, such as speech, gaze, robot behavior, or task events.

\section{Proof of Concept}
\begin{figure*}
\centering
\includegraphics[width=0.9\linewidth]{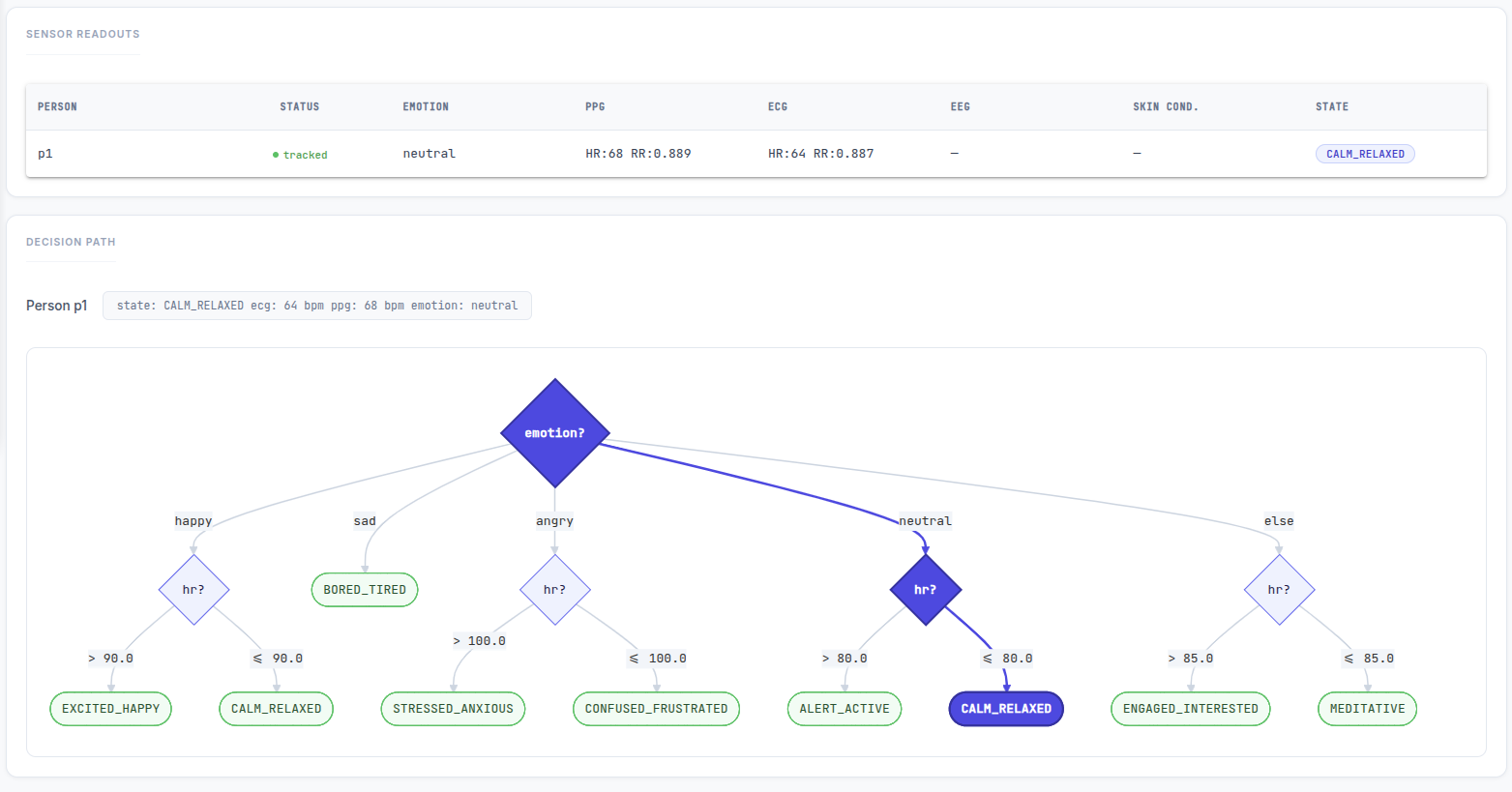}
\caption{Web GUI for real-time physiological sensor data visualization.}
\label{fig:gui}
\end{figure*}

As a proof of concept, a web-based graphical interface is implemented to visualize sensor readings and provide an accessible means of monitoring real-time physiological data during operation, as shown in Fig.~\ref{fig:gui}.

As an explanatory setup, separate sensor driver and interface nodes employing the design principles established in Section~\ref{sec:design} are developed to interface with the Polar Verity Sense PPG sensor and the Polar H10 ECG sensor. Both sensors provide raw signal data, RR intervals, and heart rate estimates over Bluetooth. Following the decoupled architecture, sensor driver nodes handle Bluetooth Low Energy communication and raw frame forwarding exclusively. Features reported by the sensor itself, in the case of the Polar sensors, RR interval and heart rate, are published on a dedicated topic, enabling downstream sensor interface nodes to compare sensor-reported features against computed features or to use them directly when computational resources are constrained. The interface nodes compute derived features including HRV metrics such as RMSSD, SDNN, and pNN50, and publish them on a dedicated feature topic.

To demonstrate a practical use case of fusing multiple sensor modalities, a simple decision tree is implemented, which estimates user affective states from the combined sensor inputs. The application makes use of the camera image stream and the \texttt{hri\_face\_detect} and \texttt{hri\_emotion\_recognition} packages from ROS4HRI, providing person detection and facial expression recognition. 

The decision tree first branches on the recognized facial expression and subsequently on heart rate thresholds derived from the averaged ECG and PPG pipelines. In this way, the demonstrator yields a more differentiated illustrative state representation than facial expression recognition alone, i.e., derived states are \textit{calm\_relaxed}, \textit{alert\_active}, or \textit{stressed\_anxious}.

This example application illustrates how multiple physiological sensors can be combined through Sense4HRI while remaining compatible with ROS4HRI perception components. It further shows how the standardized topic structure introduced in this work supports such integration without tight coupling between components.

At the current stage, the mapping between the detected person and the sensor wearer is performed statically, which limits the setup to single-user scenarios in which the person visible in the camera frame is assumed to be the individual whose physiological data is being recorded. Extending the system to multi-user operation would require a dynamic association mechanism between tracked person identities and active sensor streams, which is identified as a direction for future work. One possible solution to this challenge is to use Received Singal Stength Indicator (RSSI)-based ranging to estimate the distance to the various sensors in conjunction with the visually perceived distance.

\section{Conclusion}
This paper presents Sense4HRI, a ROS 2 extension based on ROS4HRI for integrating physiological sensors and their interpretation into human-robot interaction systems. The framework introduces a modular architecture that separates raw sensor acquisition from signal processing, reusable message interfaces for timestamped physiological data, and a topic structure compatible with ROS4HRI-based multi-user scenarios. It also supports synchronized logging of physiological signals alongside other ROS streams, making multimodal recordings easier to analyze.
The proof of concept shows how physiological sensors can be integrated into a ROS 2 pipeline and combined with other perception components relevant to HRI. The current implementation is a starting point rather than a finished solution. The demonstrator covers only a small number of sensors and relies on a static assignment between sensor data and the observed user.
Future work will focus on adding support for further physiological modalities, improving synchronization and metadata handling across devices, and enabling more flexible sensor-to-user assignment in multi-user settings.
To further increase synchronization between nodes, an event-based handshake can be implemented to coordinate message publication and processing across the system.
More broadly, the framework is meant to serve as a reusable foundation for online feature extraction, multimodal fusion, and user state modeling in ROS 2-based HRI experiments. In the longer term, such capabilities can support HDT representations for adaptive and personalized HRI.
\bibliographystyle{IEEEtran}
\bibliography{sample}
\end{document}